\documentclass[preprint,12pt]{elsarticle}

\usepackage{amssymb}
\usepackage{amsmath}
\usepackage{parskip}
\usepackage{multirow}

\usepackage{pdfpages}
\usepackage{textcomp}
\usepackage{graphicx}
\usepackage{graphics}
\usepackage {framed}
\usepackage {amsxtra}
\usepackage {enumerate}
\usepackage{commath}
\usepackage[margin=1 in]{geometry}
\usepackage{float}

\journal{arXiv}

\begin{document}

\begin{frontmatter}



\title{A simulations approach for meta-analysis of genetic association studies based on additive genetic model}


\author{Majnu John$^{a,}$$^{b,}$$^{c,}$\footnote{Corresponding author: $350$ Community Drive, Manhasset, NY 11030.  e-mail: {\sf mjohn5@northwell.edu}, Phone: +01\,718\,470\,8221,
     Fax: +01\,718\,343\,1659}, Todd Lencz$^{a,}$$^{b,}$$^{d}$, Anil K Malhotra$^{a,}$$^{b,}$$^{d}$, Christoph U Correll$^{a,}$$^{b,}$$^{d}$, Jian-Ping Zhang$^{a,}$$^{b,}$$^{d}$}

\address{$^{a}$Center for Psychiatric Neuroscience,\\ The Feinstein Institute of Medical Research,\\Manhasset, NY.}
\address{$^{b}$Psychiatry Research, Zucker Hillside Hospital, \\Northwell Health System, \\Glen Oaks, NY.}
\address{$^{c}$Department of Mathematics, \\Hofstra University, \\Hempstead, NY.}
\address{$^{d}$Departments of Psychiatry and of Molecular Medicine, \\Hofstra Northwell School of Medicine, \\Hempstead, NY.}

\begin{abstract}
  Genetic association studies are becoming an important component of medical research. To cite one instance, pharmacogenomics which is gaining prominence as a useful tool for personalized medicine is heavily reliant on results from genetic association studies. Meta-analysis of genetic association studies is being increasingly used to assess phenotypic differences between genotype groups. When the underlying genetic model is assumed to be dominant or recessive, assessing the phenotype differences based on summary statistics, reported for individual studies in a meta-analysis, is a valid strategy. However, when the genetic model is additive, a similar strategy based on summary statistics will lead to biased results. This fact about the additive model is one of the things that we establish in this paper, using simulations. The main goal of this paper is to present an alternate strategy for the additive model based on simulating data for the individual studies. We show that the alternate strategy is far superior to the strategy based on summary statistics.

\end{abstract}


\end{frontmatter}





\noindent
\section{Introduction}
 Over the last decade, genetic association studies (-both candidate based designs and genome-wide designs) have become one of the cornerstone techniques in detecting specific genetic variants related to any phenotype of interest. In association studies, genetic variation is typically measured using genotyping based on single nucleotide polymorphisms (SNPs). In order to fix ideas, let us assume that a SNP of interest is named $rs1234$ and the corresponding genotypes AA, AB and BB, with A as the major allele and B as the minor allele (the risk allele). The phenotype of interest in our motivating example-studies was weight gain due to antipsychotic treatment. With the proliferation of association studies, it is often seen that the standardized differences in weight gain (``effect sizes") between any pair of genotypes vary across the studies. For this reason, meta-analysis has become the method that is being increasingly employed to assess the phenotypic differences across the genotypic groups.

 Combining the effect sizes between genotype groups in a meta-analysis depends on the underlying genetic model. The most commonly considered genetic models are dominant, recessive and additive models. In a dominant model, allele B increases risk and the number of copies of B doesn't matter; that is, genotypes AB and BB carry the same risk. In the recessive model, two copies of allele B are required for increased risk; that is, AB is grouped together with AA, and both groups are assumed to have no risk. In an additive model, the increase in risk is proportional to the number of copies of allele B; that is, if the risk is increased $r$-fold for the genotype group AB, then it is increased $2r$-fold for BB$^{1}$.

 Typically, in the association studies considered for the meta-analysis, summary statistics (mean, median, standard deviation etc.) for weight gain will be reported for each of the three genotype groups. If we consider the underlying model to be dominant, one may take the average of the mean (or median) weight gains reported for the groups AB and BB, and calculate the standardized difference between this average and the corresponding mean (or median) weight gain reported for AA, to get the effect size for each study. It is ideal to weight the average for AB and BB by their respective sample sizes. This is a valid strategy in the sense that if the meta-analyst had access to each study's data, and if he or she grouped together the weight gains for the individual patients in groups AB and BB, and took the standardized mean difference with the individual data from AA group, then the effect size obtained will be same as the one obtained using summary statistics mentioned above; at least conceptually same - practically there might be some differences, for example, if one did not weight by sample sizes. The same argument applies for the recessive model too. However, for the additive model, the results from crude approach based on combining summary statistics will not correspond with the results using the approach based on the individual data, if the analyst was to have access to the individual data. The crude approach for the additive model would be roughly as follows. Take the difference of the summary statistics (mean or median) for AB and BB, then do similarly for AA and AB, and then take the average of these two standardized pairwise differences to obtain the corresponding effect size. One of the things that we show in this paper, via simulations, is that this crude approach leads to biased results.

 Since the approach based on summary statistics leads to biased results for the additive model, the analyst will have to fall back on subject level data from each study to calculate the appropriate effect size. But, it is well known that very rarely does a meta-analyst have access to such data. The solution to this dilemma, that we suggest in this paper, is to generate subject level data via simulations based on the summary level data, in order to calculate the effect size for each study under the additive genetic model. We show via extensive simulations and real data that the simulations-based-approach that we propose is far superior to the crude approach, and gives results very similar to those that might be obtained if individual level data was available. Before causing any confusion, we would like to clarify the word `simulations' used twice in the previous sentence. The new method proposed in this paper is based on simulations, and we compare this new method with the old method (that is, the crude approach based on summary statistics) also using simulations.

 In section 2, we explain the methods more clearly using notations. In section 3, we explain the simulations comparing the two methods, and in section 4, we present the results of the simulations. Section 5 illustrates how well the simulations-based-method works for a real data example; that is, for a data set consisting of studies for which we had subject level data. In sections 2 to 5, the phenotype of interest (e.g. weight gain) was considered as a continuous variable. But, phenotypes could be dichotomous as well (e.g. overweight - yes/no). In section 6, we show how our approach can be extended for dichotomous phenotypes as well. Finally, we summarize our findings in the last section.

\section{Methods}

In this section we describe two methods for meta-analysis assuming additive genetic model in genetic association studies. We are interested in the additive effect of the 3 genotypic groups on a phenotype of interest. In our motivating example$^{2}$, the phenotype of interest was the change in body weight due to antipsychotic treatment in patients with schizophrenia. We assume that the means $(m_{1}, m_{2}, m_{3})$ and standard deviations $(sd_1, sd_2, sd_3)$ of the phenotype of interest are given for the three groups along with the sample sizes $(n_{1}, n_{2}, n_{3})$.

\textit{Crude Approach.} One way to find the additive effect $(\beta_{crude})$ would be to stack the three means $m_{1}, m_{2}$ and $m_{3}$ into a $3 \times 1$ column vector and regress it against a column vector of group indicators $[1,2,3]^t$. The $\beta_{crude}$ thus obtained is just the average of the pairwise mean differences $m_{2} - m_{1}$ and $m_{3} - m_{2}$. The R codes corresponding to this simple approach would be


\scriptsize

\noindent \verb"M.crude" $\leftarrow$ \verb"c(m1, m2, m3) #means" \\
\verb"G.crude" $\leftarrow$ \verb"c(1, 2, 3) #group indicator variable" \\
\verb"beta.crude" $\leftarrow$ \verb"summary(lm(M.crude~G.crude))$coeff[2,1]" \\

\normalsize

In order to calculate the standard deviation of $\beta_{crude}$ (that is, the denominator of Cohen's $d$), there is no straightforward approach. However, one could take the standard deviation for the mean difference between samples 1 and 2, $sd_{12}$, and similarly for samples 2 and 3, $sd_{23}$, and take the average of these two standard deviations, $(sd_{12} + sd_{23})/2$ as the standard deviation for $\beta_{crude}$. $sd_{12}$ and $sd_{23}$ are within group standard deviations, pooled across the corresponding groups. As mentioned by Borenstein and co-authors$^{3}$, the reason to pool is that even if the underlying population standard deviations are the same, it is unlikely that the sample estimates $sd_1$ and $sd_{2}$ (used in the calculation of $sd_{12}$) and $sd_{2}$ or $sd_{3}$ (for $sd_{23}$) will be equal. The R codes for this part are

\scriptsize

\noindent \verb"SD.crude" $\leftarrow$ \verb"c(sd1, sd2, sd3)"  \\
\verb"SS.crude" $\leftarrow$ \verb"c(n1, n2, n3)"  \\
\verb"sd12" $\leftarrow$ \verb"sqrt((((SS.crude[1]-1)*SD.crude[1]^2) + ((SS.crude[2]-1)*SD.crude[2]^2))/(SS.crude[1] + SS.crude[2] - 2))"  \\
\verb"sd23" $\leftarrow$ \verb"sqrt((((SS.crude[2]-1)*SD.crude[2]^2) + ((SS.crude[3]-1)*SD.crude[3]^2))/(SS.crude[2] + SS.crude[3] - 2))"  \\
\verb"sd.beta.crude" $\leftarrow$ \verb"mean(c(sd12, sd23))" \\

\normalsize

We use the following well-known formula$^{3}$ to calculate an approximate variance for Cohen's $d$: \[ \frac{n_{1} + n_{2}}{n_{1}n_{2}} + \frac{d^{2}}{2(n_{1} + n_{2})}. \] Note that the above formula can be used only for a pair of samples. So, we use this formula to calculate $V_{d.12.crude}$ and $V_{d.23.crude}$ separately for the sample pairs 1 and 2, and 2 and 3. Similarly the correction factor \[J = 1 - \frac{3}{4(n_{1} + n_{2} - 2) - 1}, \] for the conversion of $d$ to Hedge's $g$ is also used separately for the two pairs to get $J_{12.crude}$ and $J_{23.crude}$. Then $g$ and variance of $g$ for each pair are calculated as\[ g_{12.crude} = J_{12.crude} \times d \;\; \mathrm{and}\;\; g_{23.crude} = J_{23.crude} \times d, \] \[ V_{g.12.crude} = J_{12.crude}^{2} \times V_{d.12.crude} \;\; \mathrm{and} \;\; V_{g.23.crude} = J_{23.crude}^{2} \times V_{d.23.crude}. \] Finally, $g_{crude}$ and its variance $V_{g.crude}$ are obtained as a weighted average, \[ g_{crude} = \left( \frac{g_{12.crude}}{V_{g.12.crude}} +  \frac{g_{23.crude}}{V_{g.23.crude}} \right)/\left(\frac{1}{V_{g.12.crude}} +  \frac{1}{V_{g.23.crude}} \right). \] Note that this weighted average was suggested by Hedges and co-authors$^{4}$ for combining effect size estimates.

\textit{Simulations Approach.} Another approach to the same problem would be based on simulations. In this approach, we generate individual data from the normal distribution via simulations using, for example, the \textit{rnorm} function in R, based on the mean, standard deviation and sample size for each genotypic group, reported for each study. A group indicator variable is created with $n_1$ $1$'s,  $n_2$ $2$'s and $n_3$ $3$'s, and the simulated individual data is regressed against the indicator variable to obtain $\beta_{sim}$, the numerator of Cohen's $d$. A one-way ANOVA test comparing the individual data across the 3 genotypic groups is then conducted. The denominator of Cohen's $d$ for this approach is obtained as the square root of the ratio of between-mean-squares and the $F$ statistic. This process is repeated 10,000 times and the average of Cohen's $d$'s across all the 10,000 iterations is taken as the estimated Cohen's $d$ for this approach.  The corresponding R codes are given below.

\scriptsize

\noindent \verb"Sim.Num" $\leftarrow$ \verb"10000 ## number of iterations"  \\
\\
\verb"beta.sim" $\leftarrow$ \verb"array(, Sim.Num)"  \\
\verb"sd.sim" $\leftarrow$ \verb"array(, Sim.Num)"  \\

 \verb"       for(m in 1:Sim.Num) {" \\

\verb"             sample1" $\leftarrow$ \verb"rnorm(n1, m1, sd1)"  \\
\verb"             sample2" $\leftarrow$ \verb"rnorm(n2, m2, sd2)"  \\
\verb"             sample3" $\leftarrow$ \verb"rnorm(n3, m3, sd3)"  \\
\verb"             M" $\leftarrow$ \verb"c(sample1, sample2, sample3)"  \\
\verb"             G" $\leftarrow$ \verb"c(rep(1, n1), rep(2, n2), rep(3, n3))"  \\
\\
\verb"             beta.sim[m]" $\leftarrow$ \verb"summary(lm(M~G))$coeff[2,1]" \\
\verb"             sd.sim[m]" $\leftarrow$ \verb" sqrt(anova(lm(M~G))[1,3]/anova(lm(M~G))[1,4]) }"\\

\verb"             cohen.d.sim" $\leftarrow$ \verb" mean(beta.sim/sd.sim)" \\

\normalsize

   The calculations for variance for Cohen's $d$, the correction factor for conversion from $d$ to $g$, and then for $g$ and its variance are all done in a pairwise manner as was done for the crude approach (See above).

\section{Monte Carlo Simulations}

In order to assess the performance of the two methods described above under various scenarios, we conducted Monte Carlo simulations. By ``various scenarios" we mean the scenarios obtained by varying the following parameters: number of studies in the meta analysis, the underlying distribution from which the sample for each study was obtained, the sample sizes for each study, and the true underlying effect size for each study. The last parameter was varied by varying the means for the three genotypic groups in each study, as well the within-study standard deviation. Our goals were a) to see whether the newly proposed methods improved with larger number of studies in the meta-analysis, b) whether they improved with larger sample sizes for the genotypic groups in the studies, c) whether the performance depended on either the underlying distribution or the underlying true effect sizes. It was intuitive to hypothesize that the performance of the methods would improve with increasing the number of studies and the sample sizes. So, our questions for parts a) and b) were really how large was good enough.

The key step in the simulations-based method presented above is the generation of (simulated) individual data from normal distributions using the ``reported" means, standard deviations and sample sizes of the three genotypic groups from each study. (Note that, for the simulation analysis presented in this section, there was actually no reported means or standard deviations, but these values were generated in such a way that they matched with reported values from one the real data sets (EUFEST) discussed in section 5. See below the discussion related to \textit{mean.vec} for more details.) Thus, the first question that we considered was whether the performance of the simulations-based method was affected when the original data distribution was not normally distributed. For this purpose, we conducted Monte Carlo simulations in which the original data were sampled from 3 distributions other than normal: strongly right skewed, asymmetric bimodal and heavily kurtotic. For comparison purposes, we assessed the performance of the methods, by sampling the original data from the normal distribution also. The shapes of the probability density functions for the four distributions considered are shown in Figure 1. If $g_{(\mu, \sigma)}$ denote the density function for a normal distribution with mean $\mu$ and variance $\sigma^2$, the formulas for the densities used for our simulation study are

\[ \begin{array}{lcc}
f_{1}: & g_{(0,1)} & (\mbox{standard normal}) \\
f_{2}: & \sum\limits_{l=0}^{7} \frac{1}{8} g_{\left(3\left\{(\frac{2}{3})^{l}-1\right\}, (\frac{2}{3})^{2l} \right)} & (\mbox{strongly right skewed}) \\
f_{3}: & \frac{3}{4}g_{(0,1)} + \frac{1}{4}g_{\left(\frac{3}{2}, (\frac{1}{3})^2 \right)} & (\mbox{asymmetric bimodal}) \\
f_{4}: & \frac{2}{3}g_{(0,1)} + \frac{1}{3}g_{\left(0, \frac{1}{10}\right)} & (\mbox{heavily kurtotic})  \end{array}\]

Each of the above densities is a normal mixture. These densities correspond to the first, third, eighth and fourth densities considered in Marron and Wand (1992)$^{5}$, and have been used in simulation studies previously (for example, in John and Priebe (2007)$^{6}$).

\begin{figure}[H]
\begin{center}
\includegraphics[height=7in,width=7in,angle=0]{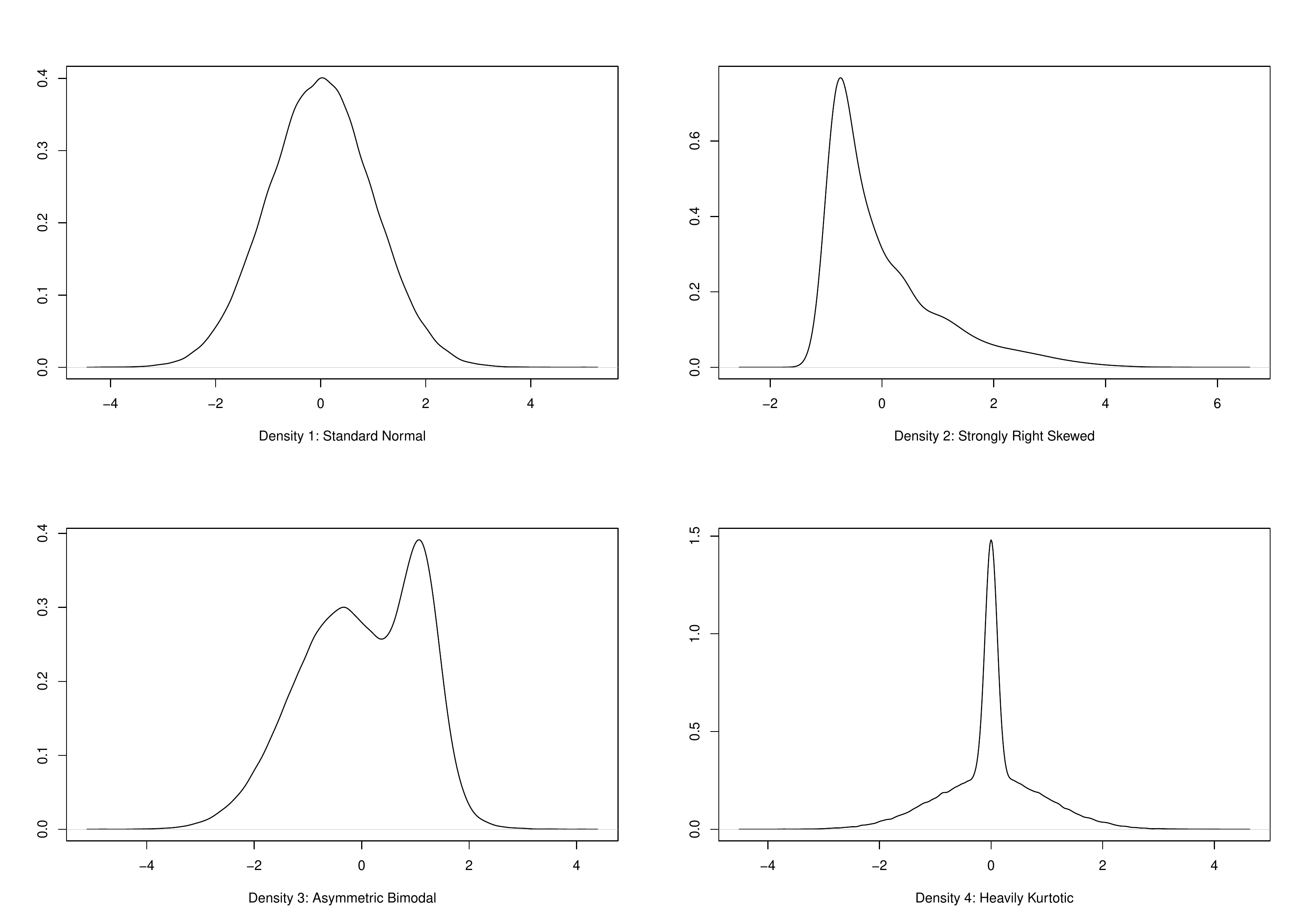}
\caption{Four densities used for simulations}
\end{center}
\end{figure}

Once we chose a distribution from among $f_1$, $f_2$, $f_3$ and $f_4$, we then selected the number of studies, $L$, to be included in the simulation analysis. We considered scenarios with $L = 5, 10$ and $15$. After the distribution and number of studies were fixed, the next task was to select a ``mean vector", \textit{mean.vec}, from among the following 3 triplets: $(4, 5.5, 7), (4, 5.5, 9)$ and $(4, 5.5, 11)$, and then to select a within-studies standard deviation $\sigma_{WS}$ to be either $\sigma_{WS} = 1$ or $5$. The approximate effect size corresponding to the various \textit{mean.vec} and $\sigma_{WS}$ are given in table 1 below.

\scriptsize

\begin{center}
\begin{tabular}{|l|c|c|c|c|}\hline
\multicolumn{5}{|c|}{Table 1. Effect size table } \\\hline
 $\sigma_{WS}$  &  $ M_{1} $  &  $ M_{2} $  &  $ M_{3} $  &  Approximate    \\
                &             &             &             &  true value of $g$        \\\hline

   \multirow{3}{*}{1}  & 4  & 5.5  & 7  & 0.82  \\\cline{2-5}
                       & 4  & 5.5  & 9  & 1.28  \\\cline{2-5}
                       & 4  & 5.5  & 11 & 1.64  \\\hline
   \multirow{3}{*}{5}  & 4  & 5.5  & 7  & 0.30  \\\cline{2-5}
                       & 4  & 5.5  & 9  & 0.48  \\\cline{2-5}
                       & 4  & 5.5  & 11 & 0.65  \\\hline
\end{tabular}
\end{center}

\normalsize

In the next step, we generated $L$ means for the $k^{th}$ genotypic group ($k = 1,2,3$) from a normal distribution with mean as the $k^{th}$ component of the \textit{mean.vec} and standard deviation equals 2, using, for example, the $rnorm$ function in R. Thus the average mean triplet across all studies will be the \textit{mean.vec} that we selected for this scenario, but we create variation across the studies by generating the mean triplet for each study from a normal distribution. Similarly, we generated $L$ (within-study) standard deviations for the $k^{th}$ genotypic group ($k = 1,2,3$) from normal distributions using the $k^{th}$ component of the $\sigma_{WS}$ selected. The codes for these steps with $L = 10$, \textit{mean.vec}$ = (4, 5.5, 9)$ and $\sigma_{WS} = 1$ are given below.

\scriptsize

\noindent \verb"L" $\leftarrow$ \verb"10 ## number of studies"  \\
\\
\verb"mean.vec" $\leftarrow$ \verb"c(4, 5.5, 9)"  \\
\verb"  m1" $\leftarrow$ \verb"rnorm(L, mean.vec[1], 2); m1[m1 < 0]" $\leftarrow$ \verb"mean.vec[1] ## L means for the 1st genotypic group" \\
\verb"  m2" $\leftarrow$ \verb"rnorm(L, mean.vec[2], 2); m2[m2 < 0]" $\leftarrow$ \verb"mean.vec[1] ## L means for the 2nd genotypic group" \\
\verb"  m3" $\leftarrow$ \verb"rnorm(L, mean.vec[3], 2); m3[m3 < 0]" $\leftarrow$ \verb"mean.vec[1] ## L means for the 3rd genotypic group" \\

\verb"sd.ws" $\leftarrow$ \verb"1"  \\
\verb"  sd1" $\leftarrow$ \verb"rnorm(L, sd, 2); sd1[sd1 < 0]" $\leftarrow$ \verb"sd.ws ## L within-study SD's for the 1st genotypic group" \\
\verb"  sd2" $\leftarrow$ \verb"rnorm(L, sd, 2); sd2[sd2 < 0]" $\leftarrow$ \verb"sd.ws ## L within-study SD's for the 2nd genotypic group" \\
\verb"  sd3" $\leftarrow$ \verb"rnorm(L, sd, 2); sd3[sd3 < 0]" $\leftarrow$ \verb"sd.ws ## L within-study SD's for the 3rd genotypic group" \\

\normalsize

With the means and standard deviations selected for the $L$ studies, we generate the three samples corresponding to the three genotypic groups for each study. The sample size considered at this step was also varied. We considered 8 sample size triplets (each triplet giving sample sizes for each of the three allelic groups), ranging from small $(10, 15, 5)$ to medium $(35, 45, 30)$ to very large $(300, 400, 240)$.

Each of the scenarios described above was repeatedly generated 500 times (that is, we used 500 Monte Carlo iterations). At each iteration, the absolute difference between the $g$ values from the original sample, and the $g$ values obtained from either of the new approaches proposed in this paper was first calculated, and the mean difference across all studies was then calculated. The mean difference thus obtained was considered as the bias for the $g$ values corresponding to either of the new methods at each iteration. The average of the above bias across all Monte Carlo iterations was presented as the mean absolute bias for $g$ in the appendix tables A1 to A4. In the above case, the bias was calculated for the $g$ for each study.

At each iteration in our Monte Carlo analysis, we also conducted a meta analysis based on random effects model to calculate a mean effect size, $g$-WM (that is, a weighted average of the $g$'s across all studies) from the original sample and a mean effect size based on the $g$'s obtained via either of the newly proposed methods. The absolute difference between the two mean effect sizes was averaged across all iterations and was labeled as the mean absolute bias for $g$-WM. Thus, across different scenarios, we compared the bias of the $g$'s for each of the individual studies as well as the bias of the weighted-averaged estimate of $g$ obtained by a meta-analysis using the random effects model.

\section{Simulation Results}

Overall, the performance of the simulations based method was better than the crude approach, in all cases that we considered. In general, the performance of the simulations based approach improved with larger number of studies and larger sample sizes for the studies, although the same cannot be said for the crude approach. The simulations based method performed fairly well even when the underlying density was strongly skewed or heavily kurtotic, but when the underlying density substantially deviated  from normal (as in the case of asymmetric bimodal density), the performance was slightly worse. The simulations based approach performed consistently well for all underlying effect sizes shown in table 1; however, the performance of the crude approach substantially worsened for larger effect sizes. We provide more details below.

\textit{Performance based on the number of studies.} As hypothesized, the performance of both the methods improved, in general, when larger number of studies were included in the simulation analyses. For example, when the original data distribution was asymmetric bimodal, with a within-study standard deviation of 1, and with the means, and the corresponding sample sizes of the three genotypic groups being $(4, 5.5, 7)$ and $(35, 45, 30)$, respectively, the bias for $g$-WM (the weighted average of effect sizes obtained using a meta-analysis based on random effects model) was $0.0661$, when the simulations based method was used and the number of studies was $5$. For the same setting, when the number of studies was increased to $10$, the corresponding bias reduced to $0.0602$ (about $8.9\%$ reduction), and when the number of studies was further increased to $15$, the bias was further reduced to $0.0479$ (about $27.5\%$ reduction). A similar trend was observed for the crude approach also: the bias values corresponding to study-sizes of $5$, $10$ and $15$ were respectively $0.2500$, $0.2492$ and $0.2479$, but the percent reductions, which amounted to $0.32\%$ and $0.84\%$ were much lower. Although the above pattern was observed in most of the scenarios there were certain cases, especially when the sample sizes were low, where the crude approach did worse with increasing number of studies. For example, when the original data distribution was strongly right skewed, with a within-study standard deviation of $1$, and with the means, and the corresponding sample sizes of the three genotypic groups being $(4, 5.5, 11)$ and $(10, 15, 5)$, respectively, the bias for $g$-WM went up from $0.6018$ to $0.7447$ and then changed to $0.7306$ as the number of studies went up from $5$ to $10$, and then to $15$, when the crude approach was used. On the other hand, for the same setting, the corresponding bias values for the simulations based approach decreased from $0.1513$ to $0.1016$ ($32.8\%$ reduction), and further to $0.0821$ ($45.7\%$ reduction) as the number of studies increased from $5$ to $10$, and then to $15$.

\textit{Performance based on the sample sizes within studies.} When all other parameters were kept the same, the bias values for both the crude approach and the simulations based approach went down, in general, as the sample size values within each study was increased. For example, when the original data was generated from the strongly right skewed density, with the three genotypic group means as $(4, 5.5, 11)$, with within-study standard deviation as $5$ and number of studies equals $10$, the mean absolute bias for $g$-WM based on the simulations approach decreased from $0.0700$ to $0.0357$ ($49\%$ reduction), and then to $0.0125$ ($82.1\%$ reduction), as the within-study sample size triplets increased from $(10, 15, 5)$ (small) to $(35, 45, 30)$ (medium), and then to $(300, 400, 240)$ (large); the corresponding bias values based on the crude approach were $0.1043$, $0.0727$ ($30.3\%$ reduction) and $0.0679$ ($34.9\%$ reduction), respectively (See appendix table A2b). However there were a few scenarios for which the bias values based on the crude approach decreased initially as the sample sizes increased from small to medium, but then increased as the sample sizes increased from medium to large, even as the bias based on the simulations based approach reduced steadily as the sample sizes increased. The following example based on appendix table A1c illustrates the above point. With the underlying data normally distributed, with the three genotypic group means as $(4, 5.5, 11)$, with within-study standard deviation as $1$ and number of studies equal to $15$, the bias values based on the crude approach based on the sample size triplets $(10, 15, 5)$, $(35, 45, 30)$ and $(300, 400, 240)$ were respectively, $0.7006$, $0.6449$ and $0.7195$ (that is, an initial reduction of $7.9\%$, but then an increase of $2.7\%$). Within the same setting and the same sample sizes, the bias based on the simulations approach decreased steadily from $0.0829$ to $0.0371$, and then to $0.0135$.

\textit{Performance based on the underlying densities.} As expected, when the underlying data distribution is normal, both the methods work better than when the data distribution is not normal. When the underlying distribution was asymmetric bimodal, the bias values of both the methods were somewhat larger compared to other distributions. When the underlying distribution was either strongly right skewed or heavily kurtotic, the bias values were slightly larger in general, yet comparable to those from the normal data distribution. For example, with a within-study standard deviation of $5$, with the means for the three genotypic groups as $(4, 5.5, 11)$, and the sample sizes as $(10, 15, 5)$, the mean absolute bias for $g$-WM, using the crude approach, for the four densities $f_1$ (normal), $f_2$ (strongly right skewed), $f_3$ (asymmetric bimodal) and $f_4$ (heavily kurtotic) are respectively $0.0935, 0.1043, 0.1567$ and $0.0894$; for the simulations based approach the corresponding four bias values were respectively $0.0692, 0.0700, 0.0953$ and $0.0707$. As mentioned above, the values under $f_2$ and $f_4$ were similar to those for $f_1$, but values for $f_3$ were much larger. The pattern remained the same even for larger sample sizes. With all the parameters exactly same as in the above example, but with sample sizes $(300, 400, 240)$, the four bias values corresponding to the densities $f_1$ to $f_4$ for the crude approach were $0.0714, 0.0679, 0.1305$ and $0.0669$, and for the simulations based approach were $0.0111, 0.0125, 0.0541$ and $0.0099$.

\textit{Performance based on effect sizes.} For small effect sizes, although the simulations based method edged out slightly over the crude approach, the performance of both the methods were more or less the same. For example, when the underlying density was strongly right skewed and number of studies was $10$ (appendix table A2b), with a within-study standard deviation of $5$ and means $(4, 5.5, 7)$ (that is, an effect size approximately $0.30$; see table 1), when the sample size triplet was $(10, 15, 5)$, the bias value for the crude approach was $0.0723$ and for the simulations based approach was $0.0684$. As the sample size increased for the same underlying effect size, the performance of both the methods improved substantially, even though the simulations based method still edged out a little bit - for the same parameters as above but with the sample size triplet as $(300, 400, 240)$, the corresponding bias values were $0.0225$ and $0.0117$. As the underlying effect size increased, the performance of the crude approach got markedly worse than that of the simulations based approach, as seen by the examples that follow. When the underlying density was strongly right skewed and number of studies was $10$, with a within-study standard deviation of $5$ and means $(4, 5.5, 11)$ (that is, an effect size approximately $0.65$), when the sample size triplet was $(35, 45, 30)$, the bias value for the crude approach was $0.0727$ and for the simulations based approach was $0.0357$. With same underlying density, same number of studies and same sample size triplet as in the above example, but with the within-study standard deviation as $1$ and the mean triplet as $(4, 5.5, 11)$ (effect size approximately $1.64$), the corresponding bias values for the two methods were $0.7001$ and $0.0512$.

\section{Real Data Example} In order to further assess the performance of the methods proposed in this paper, we applied them to a real data example for which the subject-level data was also available for the studies included in the meta analysis. This example is part of the analyses reported by Zhang and co-authors$^{2}$. Here we include only pertinent details sufficient to illustrate the methods proposed in the present paper and assess their performances. More details can be found in Zhang \textit{et al}'s paper$^{2}$. The goal of meta-analysis was to determine specific genetic variants associated with weight gain related to antipsychotic drugs. Primary outcome was change in body mass index (BMI). Literature search with inclusion/exclusion criteria narrowed down $72$ reports from $46$ independent samples. For $3$ among these $46$ independent cohorts, patient-level data were available. The three cohorts were the Second-Generation Antipsychotic Treatment Indications, Effectiveness and Tolerablity in Youth (SATIETY) study, European First Episode Schizophrenia Trial (EUFEST), and Zucker Hillside Hospital First Episode Schizophrenia Trial (ZHH-FE). Since patient-level data were available for these three cohorts, we could directly assess the additive effect of the alleles and compare it with the additive effects obtained from the crude and simulations based approaches (proposed in this paper) which utilizes reported summary means, standard deviations and sample sizes only.  In the analysis reported in Zhang \textit{et al}, SNP $rs1800544$ in the ADRA2A (adrenoceptor alpha 2A) gene was significantly associated with antipsychotic drug induced weight gain across 6 studies, including the 3 studies mentioned above, with the G allele increasing the risk. The SNP is located in the upstream of ADRA2A, and may be a binding site for transcription factors. The reported summary data for the three studies mentioned above, for the allelic groups for the SNP $rs1800544$ are presented in the table 2 below; the subscripts 1, 2 and 3 correspond to the genotypes $CC$, $CG$ and $GG$.

\scriptsize

\begin{center}
\begin{tabular}{|l|c|c|c|c|c|c|c|c|c|}\hline
\multicolumn{10}{|c|}{Table 2. Reported Summary Statistics for the 3 cohorts in the real data example } \\\hline
 Study/Cohort  &  $ M_{1} $  &  $ M_{2} $  &  $ M_{3} $  & $ SD_{1} $  &  $ SD_{2} $  &  $ SD_{3} $ &
                    $ SS_{1} $  &  $ SS_{2} $  & $ SS_{3} $  \\\hline

   SATIETY  & 11.45  & 12.16  & 14.73  & 8.29 & 8.38 & 9.63 & 63 & 63 & 42   \\\hline
   EUFEST  & 4.04  & 5.35  & 4.67  & 5.11 & 5.88 & 6.44 & 74 & 40 & 9   \\\hline
   ZHH-FE  & 3.24  & 2.44  & 3.64  & 2.11 & 1.23 & 2.42 & 25 & 24 & 21   \\\hline

\end{tabular}
\end{center}

\normalsize

 Table 3 below shows the $\beta$'s (numerator of Cohen's $d$), standard deviation of the $\beta$'s (denominator of Cohen's $d$) and the Cohen's $d$'s based on the patient level data, the crude approach and the simulations based approach. The values obtained using the simulations approach match very closely with the values calculated using patient level data, while as the values based on crude approach differed substantially. All the $\beta$'s and all the Cohen's $d$'s from the simulations approach differed from those from the patient-level data only by $0.01$ or less (percent error ranging from $0.54\%$ to $1.79\%$). On the other hand, the percent error for crude approach ranged from $4.47\%$ to $58.39\%$.

\scriptsize

\begin{center}
\begin{tabular}{|l|c|c|c|c|c|c|c|c|c|}\hline
\multicolumn{10}{|c|}{Table 3. Effect Sizes based on patient-level data and the two methods proposed in the paper } \\\hline
  & \multicolumn{3}{|c|}{Patient Level Data}   & \multicolumn{3}{|c|}{Crude Approach} &  \multicolumn{3}{|c|}{Simulations  Approach}    \\\hline
 Study/Cohort    & $\beta$ & $SD_{\beta}$ & $d$ & $\beta$ & $SD_{\beta}$ & $d$ & $\beta$ & $SD_{\beta}$ & $d$\\\hline

   SATIETY & 1.552 & 8.661 & 0.179 &  1.625 & 8.675 & 0.187 & 1.563 & 8.680 & 0.180 \\\hline
   EUFEST  & 0.746 & 5.460 & 0.137 &  0.313 & 5.470 & 0.057 & 0.742 & 5.474 & 0.136 \\\hline
   ZHH-FE  & 0.168 & 2.009 & 0.084 &  0.199 & 1.965 & 0.101 & 0.171 & 2.009 & 0.085  \\\hline

\end{tabular}
\end{center}

\normalsize

\section{The case when the phenotype of interest is a binary variable.}

The methods that we have discussed so far are applicable only when the phenotype of interest is a continuous variable, like for example, change in body weight. Sometimes, the phenotype of interest is a binary variable, and in such cases odds ratios or relative risks comparing the genotypic groups are reported. That is, if we denote the genotypic groups as, say, AA, AB and BB, then usually an odds ratio for the dichotomous phenotype of interest comparing AB and AA, and another comparing BB and AB are reported, and typically these two odds ratios differ from each other. Now the question is, if we do a meta-analysis with additive genetic model as the underlying model, how do we combine the pair of odds ratios reported within a study to get a single odds ratio for the additive model within that study. In this section, we present methods that addresses this question. Although we focus on odds ratios to explain our method, a very similar method can be worked out for relative risk as well. The method that we present here is based on a simple adaptation of the reconstruction of four-fold cell frequencies for meta-analysis by Di Pietrantonj$^{7}$. This reconstruction was further studied in Veroniki and co-authors{8}.

An overview of our adaptation is easy to describe:

\textit{Step 1}): Within each study, reconstruct the $2 \times 2$ table for each odds ratio using Di Pietrantonj's method$^{7}$. Thus, after reconstruction, with our above notation, there will be a $2 \times 2$ table comparing the dichotomous phenotype of interest between AB and AA, and another one comparing BB and AB. Note that Di Pietrantonj's method can possibly give rise to two  $2 \times 2$ tables for each odds ratio. Di Pietrantonj$^{7}$, and Veroniki (and co-authors)$^{8}$ select the correct one among these two, for each odds ratio, using a cut-off based on  the event rate in the "treatment group or the control group" (- in our case, this would be event rate in either of the genotype group). Since, in genetic association studies, this event rate is rarely reported, we use a different approach to select the correct $2 \times 2$ table for each odds ratio. More details are provided further down.

\textit{Step 2}): Once the $2 \times 2$ tables, one for each odds ratio within a study, are selected, we merge them to get a $3 \times 2$ table. That is, three rows, one for each genotypic group, and two columns, one for each category of the dichotomous variable.

\textit{Step 3}): Generate (via simulations) two variables - 1) the binary variable for the phenotype of interest and 2) a categorical indicator variable for the three genotype groups - that reflect exactly the cell frequencies in the $3 \times 2$ table obtained in step 2 above.

\textit{Step 4}): Run a logistic regression with (the \textit{logit} of) the simulated binary outcome variable as the dependent variable and the simulated 3-genotype-groups indicator as the independent variable. Exponentiate the $\beta$ co-efficient for the independent variable to get the combined odds ratio for the genetic additive model. Standard errors for the $\beta$ can be utilized to calculated the $95\%$ confidence intervals for the combined odds ratio.

  We explain and illustrate the details of the above approach using an example. Before we get to the details of our approach, we describe the example first. We generated sample data for each of the three genotype groups (AA, AB, BB) with sample size 30 in each group, that eventually yielded the following $2 \times 2$ tables:

  \begin{center}
  \begin{tabular}{ | l || c | r | }
    \hline
    Table 4a.  & phenotype present & phenotype absent \\ \hline \hline
    AB & 18 & 12 \\ \hline
    AA & 10 & 20 \\
    \hline
  \end{tabular}
\end{center}

\begin{center}
  \begin{tabular}{ | l || c | r | }
    \hline
    Table. 4b.  & phenotype present & phenotype absent \\ \hline \hline
    BB & 18 & 12 \\ \hline
    AB & 18 & 12 \\
    \hline
  \end{tabular}
\end{center}

  Details of how we generated the data and how we obtained the above $2 \times 2$ tables from the data are somewhat irrelevant to our discussion, but a reader interested in these can find the details in the appendix B.

  Thus, in our example, the odds ratio comparing AB with AA is $(18/12)/(10/20) = 3$ and the odds ratio comparing BB with AB is $(18/12)/(18/12) = 1$. If we use the formula \[ \sqrt{\left(\frac{1}{a} + \frac{1}{b} + \frac{1}{c} + \frac{1}{d}\right)} \] for the standard error of the log-odds-ratio (here, $a, b, c$ and $d$ denote the cell frequencies of a generic $2 \times 2$ table), then we get the corresponding $95\%$ confidence intervals for the two odds ratios as $(1.05, 8.60)$ and $(0.36, 2.81)$. Typically, the above odds ratios and confidence intervals are the only information reported when the results from the study's analysis are published (and hence the only information available for the meta-analyst). The $2 \times 2$ tables shown in Table 4 above are typically not reported.

  However, someone who has the original data for this study (that is, the data that we simulated for this example and that we pretend as from an original study) could use a logistic regression with the binary phenotype (present/absent) variable as the dependent variable and the 3-genotype-groups indicator as the independent variable and then exponentiate the corresponding $\beta$ to obtain the odds ratio with $95\%$ confidence interval for the additive genetic model as $1.73\; (1.02, 2.92)$ (or $1.727774 \;(1.021826, 2.921441)$ with more digits). If this (combined) odds ratio is available for the meta-analyst, then s/he doesn't have to look further (that is, s/he won't need the method described in this section). However, the combined odds ratio is seldom reported, but the two separate odds ratios (and their $95\%$ confidence intervals) for pairs of genotypic group comparisons are the ones that are typically reported. The method described in this section would help the meta-analyst to recover/estimate the unknown (-unknown to the meta-analyst, but not perhaps to the original data analyst-), combined odds ratio ($1.73$, in our example) and the confidence interval, from the two reported odds ratios (3.00 for AB vs. AA and 1.00 for BB vs. AB) and their corresponding confidence intervals.

  For a generic $2 \times 2$ table given below,

  \begin{center}
  \begin{tabular}{ | l || c | c || c |}
    \hline
    Generic  & phenotype & phenotype & genotype group \\
      Table  & present   &    absent & totals \\ \hline \hline
    Genotype group 2 & a & b & $\mathrm{m}_{1}$ \\ \hline
    Genotype group 1 & c & d & $\mathrm{m}_{2}$ \\ \hline
  \end{tabular}
\end{center}

  Di Pietrantonj's formula for recovering $a$ is \[ a = - \frac{\lambda \pm \sqrt{\lambda^2 - 4 \alpha \gamma }}{2\alpha}, \] where
  \[ \alpha = \left[(1-\mathrm{OR})^2 + \mathrm{OR}m_{2}\hat{\mathrm{SE}}^2_{\mathrm{ln(OR)}}\right] \]
  \[ \lambda = \mathrm{OR}m_{1}\left[2(1-\mathrm{OR}) - m_{2}\hat{\mathrm{SE}}^2_{\mathrm{ln(OR)}} \right] \]
  \[ \gamma = \left[ \mathrm{OR}m_{1}(\mathrm{OR}m_{1} + m_{2}) \right]. \] OR in these formulas denote the odds ratio obtained from the generic $2 \times 2$ table and $\hat{\mathrm{SE}}_{\mathrm{ln(OR)}}$ is the estimate of the standard error of $\mathrm{ln(OR)}$ obtained from the upper limit, UL, and lower limit, LL of the $95\%$ confidence interval, as \[ \frac{\mathrm{ln(UL)} - \mathrm{ln(LL)}}{2 \times 1.96}. \] Based on the $a$ estimated, we can obtain $b, c$ and $d$ as

  \[b = m_{1} - a; \;\;\;\;\;\; c = \frac{am_{2}}{\mathrm{OR}m_{1} + a(1-\mathrm{OR})}; \;\;\;\;\;\; d = \frac{\mathrm{OR}m_{2}(m_{1} - a)}{\mathrm{OR}m_{1} + a(1-\mathrm{OR})}. \] Note that there are two possible solutions for $a$, leading to two possible values for $b, c$ and $d$, and hence two possible estimates for the unknown $2 \times 2$ table. Let us denote the cell entries in the two possible $2 \times 2$ tables as $[a_{1}, b_{1}, c_{1}, d_{1}]$ and $[a_{2}, b_{2}, c_{2}, d_{2}]$. As mentioned above, Di Pietrantonj and later Veroniki and co-authors devised methods based on event rates for one of the marginal groups that would help to decide on which among these possible $2 \times 2$ tables to be picked. Such event rates are sometimes reported for epidemiological studies or clinical trials, but rarely for genetic association studies. We devised a simple scheme, more suitable to our context, that'd help us recover the correct table.

  Recall that there are two odds ratios that we are considering - one for AB vs. AA and the other for BB vs. AB, denoted from now on as $\mathrm{OR}^{\mathrm{AB}}_{\mathrm{AA}}$ and $\mathrm{OR}^{\mathrm{BB}}_{\mathrm{AB}}$, respectively. For each of these odds ratios, Di Pietrantonj's formulas$^{7}$ give two $2 \times 2$ tables. Thus in total there are four $2 \times 2$ tables. Our scheme for selecting the correct tables is best explained  by working it out for the example that we are considering at the moment. For this example, the two possible tables for $\mathrm{OR}^{\mathrm{AB}}_{\mathrm{AA}}$ are

   \begin{center}
  \begin{tabular}{ | l || c | r | }
    \hline
    Table 5a.  & phenotype present & phenotype absent \\ \hline \hline
    AB & 18 & 12 \\ \hline
    AA & 10 & 20 \\
    \hline
  \end{tabular}
\end{center}

\begin{center}
  \begin{tabular}{ | l || c | r | }
    \hline
    Table. 5b.  & phenotype present & phenotype absent \\ \hline \hline
    AB & 20 & 10 \\ \hline
    AA & 12 & 18 \\
    \hline
  \end{tabular}
\end{center}

  and for $\mathrm{OR}^{\mathrm{BB}}_{\mathrm{AB}}$ are

   \begin{center}
  \begin{tabular}{ | l || c | r | }
    \hline
    Table 6a.  & phenotype present & phenotype absent \\ \hline \hline
    BB & 12 & 18 \\ \hline
    AB & 12 & 18 \\
    \hline
  \end{tabular}
\end{center}

\begin{center}
  \begin{tabular}{ | l || c | r | }
    \hline
    Table 6b.  & phenotype present & phenotype absent \\ \hline \hline
    BB & 18 & 12 \\ \hline
    AB & 18 & 12 \\
    \hline
  \end{tabular}
\end{center}

There are four different ways to match the tables:\\

Table 5a $\leftrightarrow$ Table 6a, \;\;  Table 5a $\leftrightarrow$ Table 6b, \;\; Table 5b $\leftrightarrow$ Table 6a, \;\; Table 5b $\leftrightarrow$ Table 6b.  \\

A careful look at all the tables will immediately reveal that among the four combinations that match a table corresponding to $\mathrm{OR}^{\mathrm{AB}}_{\mathrm{AA}}$ with one of the tables corresponding to $\mathrm{OR}^{\mathrm{BB}}_{\mathrm{AB}}$, the only one that makes sense (that is, plausible) is Table 5a $\leftrightarrow$ Table 6b, because this is the only pair where the rows corresponding to AB match. In all the other three table-matching-combinations, the estimates for the row corresponding to the AB genotype group doesn't match.

In this particular example, there was one combination (Table 5a $\leftrightarrow$ Table 6b) where the rows for AB matched exactly. But, there could be other examples, where the AB-rows do not match exactly for any of the combinations. In such cases, we can calculate the ``distance" for the corresponding AB-rows for all four table match combinations listed above (e.g. the ``distance" could be the Euclidean distance, if one may consider the pair of values for the AB-row as a 2-dimensional vector), and select the combination where this distance is the minimum. Once we decide on the table-combination based on this criteria, then we can take the column-average of the corresponding AB-rows to get the AB-row for the final $3 \times 2$ table. The distances calculated for our example is given below:

\begin{center}
  \begin{tabular}{ | l || c |  }
    \hline
    Distance table.  & Distance between the corresponding AB rows \\ \hline \hline
    Table 5a $\leftrightarrow$ Table 6a & $\sqrt{(18-12)^2 + (12-18)^2} = 8.48$ \\ \hline
    Table 5a $\leftrightarrow$ Table 6b & $\sqrt{(18-18)^2 + (12-12)^2} = 0$ \\ \hline
    Table 5b $\leftrightarrow$ Table 6a & $\sqrt{(20-12)^2 + (10-18)^2} = 11.31$ \\ \hline
    Table 5b $\leftrightarrow$ Table 6b & $\sqrt{(20-18)^2 + (10-12)^2} = 2.83$ \\ \hline
  \end{tabular}
\end{center}

Again, based on this minimum distance criterion, the combination that we will select is Table 5a $\leftrightarrow$ Table 6b. In this particular example, the AB-rows for Table 5a and Table 6b are the same, $(18, 12)$ and $(18, 12)$. Hence the average is also $(18, 12)$, so that by merging Table 5a and Table 6b, we get the final $3 \times 2$ table as

\begin{center}
  \begin{tabular}{ | l || c | r | }
    \hline
    Table 7.  & phenotype present & phenotype absent \\ \hline \hline
    BB & 18 & 12 \\ \hline
    AB & 18 & 12 \\ \hline
    AA & 10 & 20 \\ \hline
  \end{tabular}
\end{center}

This completes step 2. In the next step we generate a phenotype (present/absent) variable and a 3-genotype-group indicator variable using the following R codes:

\scriptsize

\noindent \verb"al.gp.ind"  $\leftarrow$ \verb"c(rep(1, 30), rep(2, 30), rep(3, 30))" \\
                          \verb"                # 30 is the sample size for each genotype group" \\
\verb"ev.ind"  $\leftarrow$ \verb"c(sample(c(rep(1, 10), rep(0, 20)))," \\
\verb"           sample(c(rep(1, 18), rep(0, 12))),"\\
\verb"           sample(c(rep(1, 18), rep(0, 12))))" \\

\normalsize

In the R codes above, \verb"ev.ind" is the phenotype (present/absent) variable with 1 = present and 0 = absent, and \verb"al.gp.ind" is the genotype-group-indicator variable with 1 = AA, 2 = AB and 3 = BB. This completes step 3. In the final step, we run a logistic regression with \verb"ev.ind" as the dependent binary variable and \verb"al.gp.ind" as the independent variable, and then exponentiate the $\beta$ for \verb"al.gp.ind" to get the combined odds ratio for the additive genetic model as $1.727401$. This is remarkably very close to the unknown combined odds ratio from the original study in our example, $1.727774$. The difference is less than $0.001$. Using the standard error for the $\beta$, we calculate/recover the $95\%$ confidence interval as $(1.021726, 2.920464)$, which is also remarkably close to the original interval $(1.021826, 2.921441)$. Thus our slight adaptation of Di Pietrantonj's method works very well for this example. We did not conduct extensive simulations for the method presented in this section, since Di Pietrantonj$^{7}$ and Veroniki and co-authors$^{8}$ have studied elaborately the core method.

\section{Conclusion}

Meta analysis is being increasingly used to estimate the phenotype differences between genotype groups from genetic association studies. When the underlying genetic model is dominant or recessive, summary statistics from individual studies can be combined to get the pooled estimate in the meta-analysis. However, we show via simulations that when the underlying model is additive, the pooled estimate based on summary statistics leads to biased results. Since, data from individual studies is rarely available for the meta-analyst, we recommend using simulated data based on the summary statistics. We show in this paper that the method based on such simulations leads to much improved results.

\section{Acknowledgements} Majnu John\textquotesingle s  work  was  supported  in  part  by  grants  from  the  National Institute  of  Mental  Health  for  an  Advanced  Center  for  Intervention and Services Research (P30 MH090590) and a Center for Intervention Development and Applied Research (P50 MH080173).

\newpage
\appendix

\section{Tables of simulation results}

\scriptsize
\begin{tabular}{|l|c|c|c|c|c|c|c|c|c|c|}\hline
\multicolumn{11}{|c|}{Table A1a. Original Data from standard normal density } \\\hline
   \multicolumn{7}{|c|}{Number of Studies = 5}   & \multicolumn{2}{|c|}{Crude Approach} &  \multicolumn{2}{|c|}{Simulations  Approach}    \\\hline
  $\sigma_{WS}$  &  $ M_{1} $  &  $ M_{2} $  &  $ M_{3} $  &  $ N_{1} $  &  $ N_{2} $  &  $ N_{3} $ & Mean absolute & Mean absolute & Mean absolute & Mean absolute     \\
       &          &              &              &           &            &             & bias for g     & bias for g-WM    & bias for g      & bias for g-WM     \\\hline
\multirow{24}{*}{1}    &  \multirow{8}{*}{4}  & \multirow{8}{*}{5.5} & \multirow{8}{*}{7}
               & 10  & 15   & 5    & 0.3829 & 0.2299 & 0.1995 & 0.0966 \\\cline{5-11}
    &    &  &  & 15  & 20   & 10   & 0.3510 & 0.2203 & 0.1603 & 0.0794 \\\cline{5-11}
    &    &  &  & 15  & 20   & 30   & 0.2972 & 0.1701 & 0.1272 & 0.0687 \\\cline{5-11}
    &    &  &  & 15  & 45   & 30   & 0.3706 & 0.2579 & 0.1253 & 0.0623 \\
\hline
\end{tabular}

\begin{tabular}{|l|c|c|c|c|c|c|c|c|c|c|}\hline
\multicolumn{11}{|c|}{Table A1b. Original Data from standard normal density } \\\hline
   \multicolumn{7}{|c|}{Number of Studies = 10}   & \multicolumn{2}{|c|}{Crude Approach} &  \multicolumn{2}{|c|}{Simulations  Approach}    \\\hline
  $\sigma_{WS}$  &  $ M_{1} $  &  $ M_{2} $  &  $ M_{3} $  &  $ N_{1} $  &  $ N_{2} $  &  $ N_{3} $ & Mean absolute & Mean absolute & Mean absolute & Mean absolute     \\
       &          &              &              &           &            &             & bias for g     & bias for g-WM    & bias for g      & bias for g-WM     \\\hline
\multirow{24}{*}{1}    &  \multirow{8}{*}{4}  & \multirow{8}{*}{5.5} & \multirow{8}{*}{7}
               & 10  & 15   & 5    & 0.4157 & 0.1933 & 0.2326 & 0.0764 \\\cline{5-11}
    &    &  &  & 15  & 20   & 10   & 0.3377 & 0.1787 & 0.1777 & 0.0552 \\\cline{5-11}
    &    &  &  & 15  & 20   & 30   & 0.3083 & 0.1707 & 0.1342 & 0.0474 \\\cline{5-11}
    &    &  &  & 15  & 45   & 30   & 0.3551 & 0.2128 & 0.1297 & 0.0462 \\
\hline
\end{tabular}

\begin{tabular}{|l|c|c|c|c|c|c|c|c|c|c|}\hline
\multicolumn{11}{|c|}{Table A1c. Original Data from standard normal density } \\\hline
   \multicolumn{7}{|c|}{Number of Studies = 15}   & \multicolumn{2}{|c|}{Crude Approach} &  \multicolumn{2}{|c|}{Simulations  Approach}    \\\hline
  $\sigma_{WS}$  &  $ M_{1} $  &  $ M_{2} $  &  $ M_{3} $  &  $ N_{1} $  &  $ N_{2} $  &  $ N_{3} $ & Mean absolute & Mean absolute & Mean absolute & Mean absolute     \\
       &          &              &              &           &            &             & bias for g     & bias for g-WM    & bias for g      & bias for g-WM     \\\hline
\multirow{24}{*}{1}    &  \multirow{8}{*}{4}  & \multirow{8}{*}{5.5} & \multirow{8}{*}{7}
               & 10  & 15   & 5    & 0.4404 & 0.1842 & 0.2143 & 0.0738 \\\cline{5-11}
    &    &  &  & 15  & 20   & 10   & 0.3529 & 0.1937 & 0.1619 & 0.0369 \\\cline{5-11}
    &    &  &  & 15  & 20   & 30   & 0.3204 & 0.1837 & 0.1329 & 0.0329 \\\cline{5-11}
    &    &  &  & 15  & 45   & 30   & 0.3502 & 0.2064 & 0.1380 & 0.0337 \\
\hline
\end{tabular}

\begin{tabular}{|l|c|c|c|c|c|c|c|c|c|c|}\hline
\multicolumn{11}{|c|}{Table A2a. Original Data from strongly right skewed density } \\\hline
   \multicolumn{7}{|c|}{Number of Studies = 5}   & \multicolumn{2}{|c|}{Crude Approach} &  \multicolumn{2}{|c|}{Simulations  Approach}    \\\hline
  $\sigma_{WS}$  &  $ M_{1} $  &  $ M_{2} $  &  $ M_{3} $  &  $ N_{1} $  &  $ N_{2} $  &  $ N_{3} $ & Mean absolute & Mean absolute & Mean absolute & Mean absolute     \\
       &          &              &              &           &            &             & bias for g     & bias for g-WM    & bias for g      & bias for g-WM     \\\hline
\multirow{24}{*}{1}    &  \multirow{8}{*}{4}  & \multirow{8}{*}{5.5} & \multirow{8}{*}{7}
               & 10  & 15   & 5    & 0.3808 & 0.2275 & 0.2511 & 0.1179 \\\cline{5-11}
    &    &  &  & 15  & 20   & 10   & 0.3704 & 0.2077 & 0.1987 & 0.0935 \\\cline{5-11}
    &    &  &  & 15  & 20   & 30   & 0.2933 & 0.1860 & 0.1419 & 0.0744 \\\cline{5-11}
    &    &  &  & 15  & 45   & 30   & 0.4627 & 0.3399 & 0.1297 & 0.0640 \\
\hline
\end{tabular}

\begin{tabular}{|l|c|c|c|c|c|c|c|c|c|c|}\hline
\multicolumn{11}{|c|}{Table A2b. Original Data from strongly right skewed density } \\\hline
   \multicolumn{7}{|c|}{Number of Studies = 10}   & \multicolumn{2}{|c|}{Crude Approach} &  \multicolumn{2}{|c|}{Simulations  Approach}    \\\hline
  $\sigma_{WS}$  &  $ M_{1} $  &  $ M_{2} $  &  $ M_{3} $  &  $ N_{1} $  &  $ N_{2} $  &  $ N_{3} $ & Mean absolute & Mean absolute & Mean absolute & Mean absolute     \\
       &          &              &              &           &            &             & bias for g     & bias for g-WM    & bias for g      & bias for g-WM     \\\hline
\multirow{24}{*}{1}    &  \multirow{8}{*}{4}  & \multirow{8}{*}{5.5} & \multirow{8}{*}{7}
               & 10  & 15   & 5    & 0.4718 & 0.2340 & 0.2556 & 0.0832 \\\cline{5-11}
    &    &  &  & 15  & 20   & 10   & 0.3490 & 0.2008 & 0.1856 & 0.0631 \\\cline{5-11}
    &    &  &  & 15  & 20   & 30   & 0.3304 & 0.1959 & 0.1436 & 0.0450 \\\cline{5-11}
    &    &  &  & 15  & 45   & 30   & 0.4147 & 0.2617 & 0.1481 & 0.0520 \\
\hline
\end{tabular}

\begin{tabular}{|l|c|c|c|c|c|c|c|c|c|c|}\hline
\multicolumn{11}{|c|}{Table A2c. Original Data from strongly right skewed density } \\\hline
   \multicolumn{7}{|c|}{Number of Studies = 15}   & \multicolumn{2}{|c|}{Crude Approach} &  \multicolumn{2}{|c|}{Simulations  Approach}    \\\hline
  $\sigma_{WS}$  &  $ M_{1} $  &  $ M_{2} $  &  $ M_{3} $  &  $ N_{1} $  &  $ N_{2} $  &  $ N_{3} $ & Mean absolute & Mean absolute & Mean absolute & Mean absolute     \\
       &          &              &              &           &            &             & bias for g     & bias for g-WM    & bias for g      & bias for g-WM     \\\hline
\multirow{24}{*}{1}    &  \multirow{8}{*}{4}  & \multirow{8}{*}{5.5} & \multirow{8}{*}{7}
               & 10  & 15   & 5    & 0.4125 & 0.2003 & 0.2456 & 0.0814 \\\cline{5-11}
    &    &  &  & 15  & 20   & 10   & 0.3673 & 0.1902 & 0.1871 & 0.0441 \\\cline{5-11}
    &    &  &  & 15  & 20   & 30   & 0.3131 & 0.1634 & 0.1491 & 0.0477 \\\cline{5-11}
    &    &  &  & 15  & 45   & 30   & 0.4248 & 0.2665 & 0.1403 & 0.0443 \\
\hline
\end{tabular}

\begin{tabular}{|l|c|c|c|c|c|c|c|c|c|c|}\hline
\multicolumn{11}{|c|}{Table A3a. Original Data from asymmetric bimodal density } \\\hline
   \multicolumn{7}{|c|}{Number of Studies = 5}   & \multicolumn{2}{|c|}{Crude Approach} &  \multicolumn{2}{|c|}{Simulations  Approach}    \\\hline
  $\sigma_{WS}$  &  $ M_{1} $  &  $ M_{2} $  &  $ M_{3} $  &  $ N_{1} $  &  $ N_{2} $  &  $ N_{3} $ & Mean absolute & Mean absolute & Mean absolute & Mean absolute     \\
       &          &              &              &           &            &             & bias for g     & bias for g-WM    & bias for g      & bias for g-WM     \\\hline
\multirow{24}{*}{1}    &  \multirow{8}{*}{4}  & \multirow{8}{*}{5.5} & \multirow{8}{*}{7}
               & 10  & 15   & 5    & 0.4474 & 0.2842 & 0.2214 & 0.1132 \\\cline{5-11}
    &    &  &  & 15  & 20   & 10   & 0.3959 & 0.2646 & 0.1746 & 0.0903 \\\cline{5-11}
    &    &  &  & 15  & 20   & 30   & 0.3257 & 0.2465 & 0.1462 & 0.0800 \\\cline{5-11}
    &    &  &  & 15  & 45   & 30   & 0.4514 & 0.3609 & 0.1366 & 0.0744 \\
\hline
\end{tabular}

\begin{tabular}{|l|c|c|c|c|c|c|c|c|c|c|}\hline
\multicolumn{11}{|c|}{Table A3b. Original Data from asymmetric bimodal density } \\\hline
   \multicolumn{7}{|c|}{Number of Studies = 10}   & \multicolumn{2}{|c|}{Crude Approach} &  \multicolumn{2}{|c|}{Simulations  Approach}    \\\hline
  $\sigma_{WS}$  &  $ M_{1} $  &  $ M_{2} $  &  $ M_{3} $  &  $ N_{1} $  &  $ N_{2} $  &  $ N_{3} $ & Mean absolute & Mean absolute & Mean absolute & Mean absolute     \\
       &          &              &              &           &            &             & bias for g     & bias for g-WM    & bias for g      & bias for g-WM     \\\hline
\multirow{24}{*}{1}    &  \multirow{8}{*}{4}  & \multirow{8}{*}{5.5} & \multirow{8}{*}{7}
               & 10  & 15   & 5    & 0.4523 & 0.2457 & 0.2135 & 0.0889 \\\cline{5-11}
    &    &  &  & 15  & 20   & 10   & 0.3894 & 0.2341 & 0.1819 & 0.0701 \\\cline{5-11}
    &    &  &  & 15  & 20   & 30   & 0.3141 & 0.2013 & 0.1441 & 0.0752 \\\cline{5-11}
    &    &  &  & 15  & 45   & 30   & 0.3725 & 0.2546 & 0.1363 & 0.0581 \\
\hline
\end{tabular}

\begin{tabular}{|l|c|c|c|c|c|c|c|c|c|c|}\hline
\multicolumn{11}{|c|}{Table A3c. Original Data from asymmetric bimodal density } \\\hline
   \multicolumn{7}{|c|}{Number of Studies = 15}   & \multicolumn{2}{|c|}{Crude Approach} &  \multicolumn{2}{|c|}{Simulations  Approach}    \\\hline
  $\sigma_{WS}$  &  $ M_{1} $  &  $ M_{2} $  &  $ M_{3} $  &  $ N_{1} $  &  $ N_{2} $  &  $ N_{3} $ & Mean absolute & Mean absolute & Mean absolute & Mean absolute     \\
       &          &              &              &           &            &             & bias for g     & bias for g-WM    & bias for g      & bias for g-WM     \\\hline
\multirow{24}{*}{1}    &  \multirow{8}{*}{4}  & \multirow{8}{*}{5.5} & \multirow{8}{*}{7}
               & 10  & 15   & 5    & 0.4665 & 0.2749 & 0.2312 & 0.0802 \\\cline{5-11}
    &    &  &  & 15  & 20   & 10   & 0.3695 & 0.2537 & 0.1786 & 0.0683 \\\cline{5-11}
    &    &  &  & 15  & 20   & 30   & 0.3311 & 0.2384 & 0.1471 & 0.0652 \\\cline{5-11}
    &    &  &  & 15  & 45   & 30   & 0.4150 & 0.3044 & 0.1397 & 0.0659 \\
\hline
\end{tabular}

\begin{tabular}{|l|c|c|c|c|c|c|c|c|c|c|}\hline
\multicolumn{11}{|c|}{Table A4a. Original Data from heavily kurtotic density } \\\hline
   \multicolumn{7}{|c|}{Number of Studies = 5}   & \multicolumn{2}{|c|}{Crude Approach} &  \multicolumn{2}{|c|}{Simulations  Approach}    \\\hline
  $\sigma_{WS}$  &  $ M_{1} $  &  $ M_{2} $  &  $ M_{3} $  &  $ N_{1} $  &  $ N_{2} $  &  $ N_{3} $ & Mean absolute & Mean absolute & Mean absolute & Mean absolute     \\
       &          &              &              &           &            &             & bias for g     & bias for g-WM    & bias for g      & bias for g-WM     \\\hline
\multirow{24}{*}{1}    &  \multirow{8}{*}{4}  & \multirow{8}{*}{5.5} & \multirow{8}{*}{7}
               & 10  & 15   & 5    & 0.4617 & 0.2848 & 0.2477 & 0.1173 \\\cline{5-11}
    &    &  &  & 15  & 20   & 10   & 0.3794 & 0.2681 & 0.1833 & 0.0823 \\\cline{5-11}
    &    &  &  & 15  & 20   & 30   & 0.3278 & 0.1952 & 0.1527 & 0.0698 \\\cline{5-11}
    &    &  &  & 15  & 45   & 30   & 0.3809 & 0.2349 & 0.1434 & 0.0559 \\
\hline
\end{tabular}

\begin{tabular}{|l|c|c|c|c|c|c|c|c|c|c|}\hline
\multicolumn{11}{|c|}{Table A4b. Original Data from heavily kurtotic density } \\\hline
   \multicolumn{7}{|c|}{Number of Studies = 10}   & \multicolumn{2}{|c|}{Crude Approach} &  \multicolumn{2}{|c|}{Simulations  Approach}    \\\hline
  $\sigma_{WS}$  &  $ M_{1} $  &  $ M_{2} $  &  $ M_{3} $  &  $ N_{1} $  &  $ N_{2} $  &  $ N_{3} $ & Mean absolute & Mean absolute & Mean absolute & Mean absolute     \\
       &          &              &              &           &            &             & bias for g     & bias for g-WM    & bias for g      & bias for g-WM     \\\hline
\multirow{24}{*}{1}    &  \multirow{8}{*}{4}  & \multirow{8}{*}{5.5} & \multirow{8}{*}{7}
               & 10  & 15   & 5    & 0.4454 & 0.2020 & 0.2372 & 0.0818 \\\cline{5-11}
    &    &  &  & 15  & 20   & 10   & 0.3932 & 0.2059 & 0.1811 & 0.0519 \\\cline{5-11}
    &    &  &  & 15  & 20   & 30   & 0.3036 & 0.1659 & 0.1411 & 0.0479 \\\cline{5-11}
    &    &  &  & 15  & 45   & 30   & 0.4311 & 0.2731 & 0.1351 & 0.0440 \\
\hline
\end{tabular}

\begin{tabular}{|l|c|c|c|c|c|c|c|c|c|c|}\hline
\multicolumn{11}{|c|}{Table A4c. Original Data from heavily kurtotic density } \\\hline
   \multicolumn{7}{|c|}{Number of Studies = 15}   & \multicolumn{2}{|c|}{Crude Approach} &  \multicolumn{2}{|c|}{Simulations  Approach}    \\\hline
  $\sigma_{WS}$  &  $ M_{1} $  &  $ M_{2} $  &  $ M_{3} $  &  $ N_{1} $  &  $ N_{2} $  &  $ N_{3} $ & Mean absolute & Mean absolute & Mean absolute & Mean absolute     \\
       &          &              &              &           &            &             & bias for g     & bias for g-WM    & bias for g      & bias for g-WM     \\\hline
\multirow{24}{*}{1}    &  \multirow{8}{*}{4}  & \multirow{8}{*}{5.5} & \multirow{8}{*}{7}
               & 10  & 15   & 5    & 0.4574 & 0.2193 & 0.2552 & 0.0623 \\\cline{5-11}
    &    &  &  & 15  & 20   & 10   & 0.3874 & 0.2266 & 0.1844 & 0.0538 \\\cline{5-11}
    &    &  &  & 15  & 20   & 30   & 0.3315 & 0.1703 & 0.1528 & 0.0398 \\\cline{5-11}
    &    &  &  & 15  & 45   & 30   & 0.3858 & 0.2285 & 0.1427 & 0.0412 \\
\hline
\end{tabular}

\section{Data generation for tables 4a and 4b}

\normalsize

  Tables 4a and 4b were obtained as follows. (If one were to do simulation-analysis for section 6, one would repeat this procedure multiple times with variations in the mean triplet and standard deviation mentioned below.) First of all sample data from a normal distribution was generated for the genotype triplet (AA, AB, BB) assuming a mean triplet of $(4, 5.5, 7)$, and same standard deviation of 5 and equal sample size of 30 for each genotype group. The simulated phenotypic data thus obtained was dichotomized using a cut-off of 6; phenotype was considered to be present for values greater than 6, and absent otherwise. Two by two tables were obtained based on the frequencies of the dichotomized phenotype categories in the various genotype groups considered two at a time - AA vs. AB for table 4a and AB vs. BB for table 4b.

\end{document}